\documentstyle[12pt]{article}
\begin{document}
\title{\Large{{\bf Tunneling in two dimensional QCD}}}

\vspace{1.5cm}

\author{~\\{ Poul Olesen\footnote{Also at the 
Institute for Advanced Cycling, Blegdamsvej 19, Copenhagen, Denmark}} \\~\\
\it The Niels Bohr Institute\\ \it Blegdamsvej 17\\ \it DK-2100 Copenhagen {\O}
\\ \it Denmark}
\date{\today}
\maketitle
\vfill
\begin{abstract} 
The spectral density for two dimensional continuum QCD has a non-analytic 
behavior for a critical area. Apparently this is not reflected in the 
Wilson loops. However, we show that the existence of a critical area
is encoded in the winding Wilson loops: Although there is
no non-analyticity or phase transition in these Wilson loops, the dynamics 
of these loops consists of two 
smoothly connected domains separated by the critical area, one domain
with a confining 
behavior for large winding Wilson loops, and one (below the critical size) 
where the string tension disappears. We show that this can be interpreted 
in terms of a simple tunneling process between an ordered and a disordered
state. In view of recent results
by Narayanan and Neuberger this tunneling may also be relevant
for four dimensional QCD.

\end{abstract}
\vfill

\newpage

\section{Introduction}

Recently there appeared an interesting paper by Narayanan and 
Neuberger \cite{nar}
where four dimensional Wilson operators for QCD on a lattice were shown to 
have a finite and nontrivial continuum limit. One of the surprising results
is that the spectral density of eigenvalues of the Wilson loop
develops a gap at infinite $N$ for small size loops. This gap is closed when
the loop size is increased sufficiently. The resulting
eigenvalue distribution was compared to the corresponding density
in two dimensional continuum QCD, and the agreement between  four and
two dimensions was very good.

The spectral density was computed by Durhuus and the author \cite{durhuus}
from the Makeenko-Migdal equation. Later on the same result was obtained by 
Bassetto, Griguolo, and Vian \cite{vian} by direct summation of
winding Wilson loops. One result is that for a critical size of the
(dimensionless) area the spectral density becomes non-analytic. However,
the winding Wilson loops are perfectly analytical as emphasized by
Narayanan and Neuberger \cite{nar}, so the existence
of the critical loop size does not represent a  phase
transition in these loops. As a matter of fact, a superficial inspection of 
the Wilson loops defined on winding curves does not show any sign of
a critical area. So the question is what is the meaning of the
non-analyticity in the spectral density?

In this note we shall show that the critical area is indeed {\it encoded}  
in the Wilson loops with more than one winding. This encoding shows up as
a smooth transition from 
different types of physics. At short distances the winding loops oscillate,
and have a ``non-stringy'' behavior. For larger loops the string-like
behavior is regained. We compare this behavior successfully to 
tunneling in ordinary non-relatistic quantum mechanics. It turns out that the 
critical size 
corresponds to the turning point, and the smaller sizes are analogous to
the ``classically allowed'' region, whereas the confining larger loops
correspond to the ``classically forbidden'' region. In view of the
results obtained in ref. \cite{nar} a similar phenomenon most likely
happens in four dimensions.

We emphasize that the idea of tunneling from an intermediate state
with no confinement to a confining state is certainly not new.
Neuberger \cite{herbert} has discussed this several years ago, and
it is also mentioned in ref. \cite{nar}. Our new point is that
in contrast to what one would naively think by inspection of the winding Wilson
loops, these have encoded information on a tunneling transition with a turning
point exactly at the critical area. 

In Section 2 we summarize the results on two-dimensional $N=\infty$ QCD.
In Section 3 we develop asymptotic results for Wilson loops with
a large number of  windings, and in Section 4 we connect these
loops with the gap in the spectral density. In Section 5 we develop the
tunneling picture in the WKB approximation, and make some
conclusions.

\section{The gap and non-analyticity in the spectral density}

We start by giving an overview of the results previously obtained 
\cite{durhuus,vian} in two dimensional QCD which are relevant for the 
following. 
The spectral density gives the distribution of the continuous $N=\infty$ 
eigenvalues $\theta$. From \cite{durhuus}, eqs. (4.1) and (4.2), one
easily finds 
\begin{equation}
 \theta = \arccos \left(\cosh y-A~\frac{\sinh y}{2y}\right)+\frac{y}{\sinh y}
\sqrt{1-\cosh^2y-A^2\frac{\sinh^2y}{4y^2}+A\frac{\sinh y \cosh y}{y}},
\label{1}
\end{equation}
where $y=-\pi A\rho_A(\theta)$, and where $A=g^2N \times$(Area), which means 
that $A=2k$, where $k$ is the parameter used in \cite{durhuus}.
The parameter $A$ is a dimensionless area,
defined such that the Wilson loop has the behavior
\begin{equation}
W=e^{-A/2}.
\end{equation}
There exists a finite gap $\theta_c\leq \pi$ for $A<4$, given by
\begin{equation}
\theta_c=\sqrt{A-\frac{A^2}{4}}+\arccos (1-\frac{A}{2}).
\label{critical}
\end{equation}
For $A\geq 4$ the gap disappears, and the maximum value of $\theta$ is $\pi$.

Near the gap one has
\begin{equation}
\rho_A(\theta )\approx \frac{\sqrt{2(\theta_c-\theta)}}{\pi A(\frac{4}{A}-1)^
{1/4}},~~~~\theta\sim\theta_c.
\end{equation}
Thus the derivative $\partial\rho_A(\theta)/\partial\theta$ becomes
infinite at the gap.

Close to $A=4$ this breaks down, and instead one gets
\begin{equation}
\rho_A(\theta)\approx\frac{1}{4\pi}\left(\frac{9\sqrt{3}}{2}~(\pi-\theta )
\right)^{1/3}.
\end{equation}
For $A>4$ the behavior near $\theta =\pi$ is smooth. The value $y_c$ at
$\theta=\pi$ is determined by
\begin{equation}
\cosh y_c-A\frac{\sinh y_c}{2y_c}=-1.
\label{c}
\end{equation}
Expanding around $y=y_c$ with $y\approx y_c+\delta y$, we get
\begin{equation}
\delta y\approx (\pi-\theta)^2\left[2 \left(1-\frac{y_c}{\sinh y_c}\right)^2
\left((1+\frac{A}{2y_c^2})\sinh y_c-\frac{A}{2y_c}\cosh y_c\right)\right]^{-1}.
\label{xx}
\end{equation}
From this we see that the derivative  $\partial\rho/\partial\theta$ vanishes
at $\theta=\pi$. For $A$ large, $y_c\approx -A(1-2e^{-A/2})/2$, and one obtains
\begin{equation}
\rho_A(\theta)\approx\rho_A(\pi)+\frac{1}{2\pi}~e^{-A/2}(\pi-\theta)^2+...~.
\end{equation}

For $A\rightarrow 4+$ we can easily find the small $y_c$ from (\ref{c}),
\begin{equation}
y_c \approx -\sqrt{3 (A-4)}.
\end{equation}
Therefore the derivative $\partial\rho_A(\theta)/\partial A$ becomes infinite 
for $A\rightarrow 4$. In the following we shall show that this behavior
reflects itself in the winding Wilson loops as  tunneling with the
turning point $A=4$.

\section{The Wilson loops with windings}

The Wilson loop with $n$ windings
found by Kazakov and Kostov \cite{kazakov} is completely analytic in $A$,
\begin{equation}
W^{(n)}(A)=\frac{1}{n}~L^1_{n-1}(nA)~e^{-nA/2},
\label{wilson}
\end{equation}
where $L^1_{n-1}$ is the Laguerre polynomial of type one,
\begin{equation}
L^1_{n-1}(nA)=\sum_{k=0}^{n-1}\frac{n!n^k}{(k+1)(k!)^2(n-k-1)!}~(-A)^k.
\label{laguerre}
\end{equation}
The lack
of analyticity of $\rho$ for $A=4$ obviously occurs due to the summation 
defining the spectral density,
\begin{equation}
\rho_A(\theta)=\frac{1}{2\pi}~\left[1+2\sum_{n=1}^\infty \cos (n\theta)~
W^{(n)}(A)\right].
\label{rho}
\end{equation}
Also, from (\ref{wilson}) it appears that $W^{(n)}$ has no reference
whatsoever to the occurence of the gap $\theta_c$ in the spectral density.

The non-analyticity and the existence of a gap for $A<4$ is clearly
related to the large $n$ behavior of $W^{(n)}$, since if the sum in (\ref{rho})
is truncated, these phenomena do not occur. This is well known to
several authors (for a recent reference, see \cite{nar}). However, we shall 
now show that $W^{(n)}(A)$ has a behavior which, at least for large
$n$, reflects the existence of the gap $\theta_c$. To this end we need
the {\it uniform} large $n$ asymptotic behavior of the Laguerre polynomial
found by Erd\'elyi  \cite{erdeley} from Laguerre's differential
equation, and later also found from integral representations of
the Laguerre polynomial \cite{frenzen}. There are two (partially
overlapping) representations of these asymptotes, called the
{\it Bessel form} and the {\it Airy form}. Applying Erd\'elyi's
Bessel form to eq. (\ref{wilson}) we get
\begin{equation}
W^{(n)}(A)\approx \frac{\sqrt{2}}{nA}\left(\frac{\sqrt{t-t^2}+\arcsin\sqrt{t}}
{\sqrt{1/t-1}}\right)^{1/2}J_1\left(2 n(\sqrt{t-t^2}+\arcsin \sqrt{t})\right)
+...~,
\label{bessel}
\end{equation}
where $t=A/4$. We have checked numerically that the corrections ``...'' 
are extremely small even for moderate values of $n$, except for
$A\approx 4$, where the approximation breaks down. However, the
larger $n$ is, the closer to $A=4$ does the approximation work.

The Bessel form is thus valid for $A<4$ and $A\neq 0$.

On the other hand, applying Erd\'elyi's Airy form, we obtain for $0<t\leq 1$
\begin{equation}
W^{(n)}(A)\approx \frac{(-1)^{n-1}2^{\frac{1}{2}}3^{\frac{1}{6}}}
{n^{\frac{4}{3}}A}\left(\frac{(\arccos
\sqrt{t}-\sqrt{t-t^2})^{\frac{1}{3}}}{\sqrt{1/t-1}}\right)^{\frac{1}{2}}
Ai\left(-(3n)^{\frac{2}{3}} (
\arccos\sqrt{t}-\sqrt{t-t^2})^{\frac{2}{3}}\right).
\label{airy1}
\end{equation}
For $t\geq 1$ we get from \cite{erdeley}
\begin{equation}
W^{(n)}(A)\approx \frac{(-1)^{n-1}2^{\frac{1}{2}}3^{\frac{1}{6}}}{n^{\frac{4}
{3}}A}\frac{(\sqrt{t^2-t}-\ln (\sqrt{t}+\sqrt{t-1})^{\frac{1}{6}}}{(1-1/t)^
{\frac{1}{4}}}~Ai((3n)^{\frac{2}{3}}(\sqrt{t^2-t}-\ln (\sqrt{t}+\sqrt{t-1}))^
{\frac{2}{3}}).
\label{airy2}
\end{equation}
The Airy function is defined by
\begin{equation}
Ai(z)=\frac{1}{\pi}~\int_0^\infty \cos (\frac{1}{3}t^3+zt)dt
\end{equation}
and the Airy forms (\ref{airy1}) and (\ref{airy2}) can be expressed in terms of
Bessel functions by means of the formulas
\begin{equation}
Ai(z)=\frac{1}{3}\sqrt{|z|}~\left(J_{1/3}\left(\frac{2|z|^{3/2}}{3}\right)+
J_{-1/3}\left(\frac{2|z|^{3/2}}{3}\right)\right)~~{\rm for}~~z<0,
\label{aij}
\end{equation}
and
\begin{equation}
Ai(z)=\frac{1}{\pi}\sqrt{\frac{z}{3}}~K_{1/3}\left(\frac{2z^{3/2}}{3}\right)~~
{\rm for}~~z>0.
\label{aik}
\end{equation}
It should be noticed that these Bessel functions are different from the $J_1$
Bessel function in eq. (\ref{bessel}).

\section{Connection between the gap parameter $\theta_c$ and $W^{(n)}$}

In this Section we shall show that in spite of their appearence the
$W^{(n)}$'s with $n>1$ know about the gap, and have quite different
behavors on the two sides of $A=4$.

If we consider $A<4$ we see that the argument of the Bessel function 
(\ref{bessel}) is similar to the gap parameter $\theta_c$ given
in eq. (\ref{critical}). The connection is easily found,
\begin{eqnarray}
&&\sqrt{t-t^2}+\arcsin\sqrt{t}=\frac{1}{2}\left(\sqrt{A-\frac{A^2}{4}}+2
\arccos\sqrt{1-\frac{A}{4}}\right)\\ \nonumber
&=&  \frac{1}{2}\left(\sqrt{A-\frac{A^2}{4}}+
\arccos (1-\frac{A}{2})\right)=\frac{1}{2}~\theta_c.
\end{eqnarray}
Therefore eq. (13) can be written
\begin{equation}
W^{(n)}(A)=\frac{\theta_c^{1/2}}{nA(4/A-1)^{1/4}}~J_1(n\theta_c)+...~.
\label{besselnew}
\end{equation}
Similarly, we also have
\begin{equation}
\arccos\sqrt{t}-\sqrt{t-t^2}=\frac{1}{2}(\pi -\theta_c),
\end{equation} 
so we can rewrite eq. (\ref{airy1}),
\begin{equation}
W^{(n)}(A)\approx \frac{(-1)^{n-1}2^{1/3}3^{1/6} (\pi-\theta_c)^{1/6}}{n^{4/3}
A(4/A-1)^{1/4}}~Ai(-(\frac{3n}{2})^{2/3}(\pi-\theta_c)^{2/3}).
\end{equation}
By means of eq. (\ref{aij}) this can also be written in terms of Bessel 
functions,
\begin{equation}
W^{(n)}(A)\approx \frac{(-1)^{n-1} (\pi-\theta_c)^{1/2}}{\sqrt{3}~n
A(4/A-1)^{1/4}}\left[J_{1/3}(n(\pi -\theta_c))+J_{-1/3}
(n(\pi -\theta_c))\right].
\label{airynew}
\end{equation}
The various Bessel functions in these expressions oscillate, and the
oscillations are governed entirely by the gap parameter, either
$\theta_c$ directly as in (\ref{besselnew}), or else by the
length of the gap $\pi-\theta_c$ as in (\ref{airynew}). Thus, the
Wilson loops with windings $n>1$ actually know about the gap
parameter.

We have checked numerically that (\ref{besselnew}) and (\ref{airynew}) give
a very precise representation of the exact expression for $W^{(n)}$
in eq. (\ref{wilson}). This holds for surprisingly low values of $n$, 
especially as far as the Airy form (\ref{airynew}) is concerned. With
the Bessel form (\ref{besselnew}) there are deviations close to $A=4$,
but one can get closer by taking large $n'$s, e.g. $n=10$.

When $n$ is sufficiently large, and $\theta_c$ or $\pi -\theta_c$
are not too small, we can use the well known asymptotic representations
for the Bessel functions to rewrite (\ref{besselnew}) and (\ref{airynew}),
\begin{equation}
W^{(n)}(A)\approx \frac{\sqrt{2}}{\sqrt{\pi}~n^{3/2}A(4/A-1)^{1/4}}~\cos
(n\theta_c-\frac{3\pi}{4})+...~.
\label{cos}
\end{equation}
Although the Bessel functions are  different in eqs. (\ref{besselnew}) 
and (\ref{airynew}), after use of the addition theorem for cosines they lead 
to the same expression (\ref{cos}), as one would expect. The asymptotic
expression (\ref{cos}) can be derived in a straightforward manner
by the saddle point method from the integral representation of the Laguerre
polynomial,
\begin{equation}
L_{n-1}^1(nA)=\int_C\frac{dt}{2\pi i}~e^{-nAt}\left(1+\frac{1}{t}\right)^n,
\label{contour}
\end{equation}
where the contour $C$ encloses the origin.

The asymptotic expression (\ref{cos}) gives good results if $A$ is not
too close to 4. As an example, let us check the zeros predicted by
(\ref{cos}) against the zeros in $L_{n-1}^1(nA)$ for the rather
moderate value $n=5$. The $\theta_c(A)'$s which correspond to the zeros
in $L_4^1(5A)$ are
\begin{equation}
0.766,~~1.403,~~2.034,~~{\rm and}~~2.662
\end{equation}
The zeros predicted by (\ref{cos}) are 
\begin{equation}
0.785,~~1.414,~~2.042,~~{\rm and}~~2.670.
\end{equation}
The largest discrepancy is of order three per cent. With $n>5$ the accuracy 
improves, of course. 

We now turn to $A>4$, where we have eq. (\ref{airy2}). By use of (\ref{aik}) 
we can rewrite this expression as
\begin{eqnarray}
W^{(n)}(A)&\approx&\frac{(-1)^{n-1}\sqrt{2}}{\pi nA(1-4/A)^{1/4}}~
\left(\frac{A}{4}
\sqrt{1-\frac{4}{A}}-\ln(\frac{\sqrt{A}}{2}+
\sqrt{\frac{A}{4}-1})\right)^{1/2} \\ \nonumber
&&\times K_{1/3}(\frac{nA}{2}(\sqrt{1-\frac{4}{A}}-\frac{4}{A}\ln (
\frac{\sqrt{A}}{2}+\sqrt{\frac{A}{4}-1}~))).
\label{30}
\end{eqnarray}
By means of the asymptotic expansion of the $K-$function this gives ($A>4$)
\begin{equation}
W^{(n)}(A)\approx \frac{(-1)^{n-1}}{\sqrt{2\pi}~n^{3/2}A(1-4/A)^{1/4}}~
\left(\frac{\sqrt{A}}{2}+\sqrt{\frac{A}{4}-1}\right)^{2n}~\exp\left[-\frac{nA}
{2}\sqrt{1-\frac{4}{A}}\right].
\label{exp}
\end{equation}
This result can also be derived by the saddle point method from the integral
representation (\ref{contour}).  
For $A\rightarrow \infty$ eq. (\ref{exp}) leads to
\begin{equation}
W^{(n)}(A)\approx \frac{(-A)^{n-1}e^n}{\sqrt{2\pi}~n^{3/2}}~e^{-nA/2}.
\label{expn}
\end{equation}
This is precisely the expression one would get from the exact
formula (\ref{wilson}) by taking into account only the first term
in the Laguerre polynomial (\ref{laguerre}) and using Stirling's
formula to estimate the factorials for $n\rightarrow\infty$.

It is easy to check that eqs.(\ref{airynew}) and (\ref{exp}) give good 
numerical results
for $A\geq 4$. The same is not true for (\ref{expn}), which is
only valid for $A'$s somewhat larger than 4.

The turning point region around $A=4$ can be approximated by simpler
expressions. The results depend on how close $A$ is to 4. If
$A-4 <<n^{-2/3}$ one finds \cite{erdeley}
\begin{equation}
W^{(n)}(A)\approx \frac{(-1)^{n-1}}{(2n)^{4/3}3^{2/3}\Gamma(2/3)},
\end{equation}
whereas for $A-4<<n^{-2/5}$ the approximation is
\begin{equation}
W^{(n)}(A)\approx \frac{(-1)^{n-1}}{(2n)^{4/3}}~Ai (n^{2/3}(A-4)/(2 2^{1/3})).
\end{equation}
These two expressions agree if $A=4$ because $Ai (0)=1/(3^{2/3}\Gamma(2/3))$.

\section{On the physical meaning of the gap and  non-analyticity: Tunneling}

The results obtained in the previous section show that the Wilson loop 
$W^{(n)}$ with two or more windings is influenced by the gap and the
$A=4$ behavior. There are two different physical regions, but
they are smoothly connected: In the region $A<4$ there is no area behavior, 
but oscillations. In the second case, there is an area behavior. This is
in contrast to $W^{(1)}$, where there is always an area behavior $e^{-A/2}$.

One may wonder why the area behavior is not present for $A$ less than 4,
because the Wilson loop (\ref{wilson}) contains the factor $e^{-nA/2}$,
which has a large negative exponent for large $n$, even if $A$ is moderate. 
The answer to this point is that the Laguerre polynomial has  large
coefficients for large $n$. For example, the leading term is of order
$A^ne^n$, which for moderate $A$ can win over $e^{-nA/2}$.

The non-analytic behavior of
the spectral density must be a non-smooth reflection of the smooth 
transition between the
two physically different regions, separated by $A=4$. The Wilson loops 
$W^{(n)}$ are analytic in $A$, as follows trivially from (\ref{wilson}). 
However, if we think in terms of a string 
description of the dynamics, we see that the Wilson loop with $n>1$ is
able to go from the ``evident'' string at larger $A$ to something
which does not look ``stringy'' at smaller $A'$s. 

Let us  look at the eigenvalue gap parameter $\theta_c(A)$ as a coordinate,
which is compact but continuous (thanks to $N=\infty$) in the
region $A<4$. For $A>4$ we can instead take the argument
of the Bessel $K$-function  as a coordinate. Let us 
denote this set of ``coordinates'' by $\xi$,
\begin{equation}
\xi_1 =\pi-\theta_c(A)~~(A<4),~~\xi_2 =\sqrt{A^2/4-A}-
2\ln (\sqrt{A}/2+\sqrt{A/4-1})~~(A>4).
\label{xi}
\end{equation}
Then the (smooth) transition that occurs at $A=4$ 
looks very much like tunneling, where $A<4$ is the ``classically
allowed region'', whereas  $A>4$ is the region which is ``classically
forbidden'', and $A=4$ is the turning point. The Bessel functions we
found are indeed similar to those wave
functions that are encountered in the WKB approximation.

To see the connection to the WKB approximation in details, we notice that
$\xi$ should be represented by an integral over ``momenta'' $k_L$ and $k_R$''. 
The integrals are
\begin{equation}
\int_A^4 k_L dx=n~\xi_1 ~~{\rm  and}~~\int_4^A k_Rdx=n~\xi_2,
\label{wkb}
\end{equation}
where we consider $n$ to be a fixed number. From eqs. (\ref{xi}) we get
\begin{equation}
k_L=\frac{n}{2}~\sqrt{\frac{4-A}{A}},~~~k_R=\frac{n}{2}~\sqrt{\frac{A-4}{A}}.
\end{equation}
Near the turning point $A=4$ we therefore see that the squares of the momenta
are linear, $\propto 4-A$ or $A-4$. Therefore the WKB-solution 
on the left should be expresed in terms of $\xi_1^{1/2}J_{\pm 1/3} (n\xi_1)$ 
and to the right by $\xi_2^{1/2}I_{\pm 1/3} (n\xi_2)$ (see e.g. 
ref. \cite{schiff}, pp. 188-190). Taking into account that
\begin{equation}
K_{1/3}(x)=\frac{\pi}{\sqrt{3}}~[I_{-1/3}(x)-I_{1/3}(x)].
\end{equation}
this is precisely what we obtained using Erd\'elyi's approximations. This 
even applies to the prefactors
\begin{equation}
(4/A-1)^{-1/4}~~{\rm or}~~ (1-4/A)^{-1/4},
\end{equation}
which correspond to the prefactors $k_L^{-1/2}$ or $k_R^{-1/2}$ occuring
in the WKB expressions \cite{schiff}.

It should be noticed that even without using the large $n$ approximate results,
the Wilson loops (\ref{wilson}) for $n>1$ clearly distiguish between
values of $A$ larger or smaller than 4. This for example follows
because the Laguerre
polynomials $L_{n-1}^a(x)$ have all their zeros in the region
$0<x<\nu$, where $\nu =4n+2a-2$, and outside this interval the polynomials
are monotonic \cite{erdeley}. In our case (take $x=nA$) this condition for
oscillations means $0<A<4$, so the zeros
are all below $A=4$, and in this region the Wilson loops (\ref{wilson})
oscillate. There is therefore always a non-stringy behavior below $A=4$
for $n\geq 2$. In this connection it is amusing to notice that if
we define a ``wave function'' $\psi$
\begin{equation}
\psi_n(A)=A~W^{(n)}(A),~~n>1,
\end{equation}
then from (\ref{wilson}) and the differential equation for the Laguerre
polynomials it follows that $\psi$ satisfies the linear 
``Schr\"odinger equation''
\begin{equation}
\frac{d^2\psi_n(A)}{dA^2}+\frac{n^2}{4}\left(\frac{4}{A}-1\right)\psi_n(A)=0.
\label{schr}
\end{equation}
In this equation no approximation has been used. Eq. (\ref{schr}) clearly
shows the existence of the turning point at $A=4$. The approximate results
discussed in the preceeding sections were actually derived from 
(\ref{schr}), see ref. \cite{erdeley}.

The physics behind the domains separated by $A=4$ is indicated by
the spectral density. In the extreme case where $A$ is small, the
eigenvalues $\theta$ are centered around $\theta =0$, and in the other 
extreme where $A$ is large, the eigenvalues are almost uniformly distributed.
Thus, the domain to the right of $A=4$ corresponds approximately to 
a  completely disordered state, whereas the domain to the left of
$A=4$ is a much more ordered state, at least as far as the eigenvalues are 
concerned. Therefore the confinement is related to randomness, and
the approximately ordered state leads to an ``evaporation'' of the
effective string tension. This  evaporation happens to the
left and even somewhat to the right of of $A=4$, see the exponent in 
eq. (\ref{exp}).

The coordinates $\xi_{1,2}$ represent a new degree of freedom, which 
perhaps may be interpreted as an extra dimension, following ideas by Das and 
Jevicki \cite{das}
and by Brower and collaborators \cite{brower}. What this adds up to
is that  $\xi_{1,2}$ might play a dynamical role in an
effective string theory in three dimensions, where the action provides a 
barrier, which
produces tunneling in the new dimension. One difficulty with this picture is 
that this tunneling does not occur for $W^{(1)}$. 

\section{Conclusions}

The main conclusion of this note is that the behavior of $W^{(n)}(A)$ as
a function of $A$ physically represents  a smooth
tunneling between a short distance ($A<4$) and a large distance state ($A>4$).
In view of the results obtained by Narayanan and Neuberger \cite{nar} on
the relation between four and two dimensional QCD this may be true not
only in two, but also in four dimensions. The $A<4$ region may be considered 
as a transient state where there is no ``obvious'' string-picture. The
tunneling to the disordered confined state has undoubtedly different causes
in different dimensions (instantons, monopoles, vortex lines,...). In four 
dimensions the asymptotic freedom 
short-distance region would be hooked on to the approximately ordered
state, which is then intermediate between freedom and confinement.
Of course, such a picture does not exclude that there exists a sufficiently 
intelligent string theory
which can accomodate this behavior, as advocated by several authors,
most recently by Brower \cite{brower}. However, it is of some interest that the
$W^{(n)}$'s have at least some knowledge of the complexity involved
at different scales.

The connection between disorder and large distance dimensional reduction
was actually argued a long time ago \cite{ambjorn} to be connected to 
confinement, because
e.g. four dimensional QCD could reduce to two dimensional QCD at
large distances as a consequence of the disorder. The results obtained 
in \cite{nar} precisely means the existence of a disordered vacuum at large
distances in four dimensions. The phenomenon disorder $\rightarrow$ reduced
dimension is known to occur in simple condensed matter models \cite{condens},
e.g. a scalar field  coupled to a random Gaussian field. In this
relatively simple case the disorder reduces the effective
dimension $D$ by two, $D\rightarrow D-2$. The
complexity of QCD means that the results and the methods from
solid state physics have not been generalized to QCD, and it is not known
by how much the dimension effectively reduces. Thus it may be that both four
and three dimensions effectively reduce to two dimensions. In QCD the 
disordered field should be generated dynamically in the ``true'' vacuum, and 
not put in externally, as in the condensed matter case. The complexity of 
QCD is a good reason to study this problem by using lattice QCD. It
would be interesting to investigate on a lattice whether the tunneling
transition found in the present paper also occurs for (winding)
Wilson loops in four dimensions. The similarity of the spectral density in
four and two dimensions found in \cite{nar} indicates that this would be 
the case.

\vspace{.5cm}

{\bf Acknowledgement:}\\

I thank Poul Henrik Damgaard for valuable discussions. I also thank
Herbert Neuberger for valuable remarks which improved the presentation
of my results.

\end{document}